\documentstyle[12pt]{article}
\begin{document}
\thispagestyle{empty}

\begin{center}
\LARGE \tt \bf{A teleparallel effective geometry for Einstein's unified field theory}
\end{center}

\vspace{2.5cm}

\begin{center} {\large L.C. Garcia de Andrade \footnote{Departamento de
F\'{\i}sica Te\'{o}rica - Instituto de F\'{\i}sica Rua S\~{a}o Fco. Xavier 524, Rio de Janeiro, RJ

Maracan\~{a}, CEP:20550-003 , Brasil.

E-Mail.: garcia@dft.if.uerj.br}}
\end{center}
\vspace{2cm}

\begin{abstract} 
Riemannian and teleparallel geometrical approaches to the investigation of Maxwell electrodynamics shown that a unified field theory of gravitation and electromagnetism a la Einstein can be obtained from a stationary metric. This idea contrasts with the recently proposed pre-metric electrodynamics by Hehl and Obukhov. In the teleparallel case the definition of the electric field is obtained straightforward from the spacetime metric and the orthonormal basis frame of teleparallelism. In this case the only nonvanishing component of Cartan torsion is defined as the effective electric field. In this approach the gravitational potentials or metric coefficients are expressed in terms of the effective or analogous electric and magnetic potentials. Thefore the Maxwell equations in vacuum can be obtained by derivation of this electric field definition as usual. In the Riemannian case we consider an electrostatic spacetime where the Einstein equations in vacuum in the approximation of linear fields. The constraint of Einstein equations in vacuum are shown to lead or to the Coulomb equation or to a singular behaviour on the metric which would represent a kind of effective electrodynamic black hole event horizon.  
\end{abstract}

\newpage
\section{Introduction}
Recently Hehl and Obukhov (HO) \cite{1} have proposed a pre-metric electrodynamics where the metric would be oobtained from constitutive relations of electrodynamics. Here we consider an alternative approach to a geometrical electrodynamics where instead of obtained the metric from the constitutive electrodynamice equations we obtain the Maxwell's equation in vacuum from teleparallel theory , in much the same way Einstein and Cartan dreammed to obtained as such unified field theory \cite{2} , where the metric coefficients would be obtained from the electrodynamics potential, unifying in this way the concepts of gravitational and electromagnetic potentials. The metric is in fact is considered as a e.m perturbation of flat Minkowski spacetime. In the case of Riemannian electrostatics we show that it reduces from the Riemann-flat or Minkowskian vacuum electrodynamics in the case of linear electrodynamics. The case the Riemann tensor does not vanish we obtain a nonvacuum electrostatic where the Riemann curvature tensor would represent the density of charge. Actually we show that the Einstein equations in vacuum reduce to the trivial Minkowskian electrodynamics from the static metric representing this electrostatic spacetime. Another interesting feature of this spacetime geometry is that a singularity in the metric $g_{00}=0$ or an event horizon may appear in the theory similar to what happens in Novello et al \cite{3} black holes in flowing dielectric effective metric in Riemannian spacetime. The main difference here being that we are in vacuum and our metric is not written in terms of the electric and magnetic fields but actually in terms of electric and magnetic potentials. Besides in Novello's approach the effective metric depends on the nonlinear electric fields and not on the nonlinear e.m potentials as here. Since the generalization to higher dimensions is straightforward to simplify matters we consider that the electric and magnetic potentials are unidimensional in space. We are not proposing an alternative electrodynamics to HO  electodynamics but actually some simple and very pedestrian approach of obtained Maxwell's theory from the stationary electromagnetic and electrostatic metrics from non-Riemannian geometry. Teleparallel electrodynamics have also been proposed by V. de Andrade and J. Pereira \cite{4}, however in their theory of electrodynamics no e.m effective metric is proposed. The term effective metric \cite{5} here is used in the context that the potentials in the metric maybe not necessarilly consider to represent the e.m potentials but simply mathematical functions and yet in this case gravitational equations electromagneticeffective geometry could be consider. Note also that no attempt is made here to think on a analogy to the Einstein-Cartan theory \cite{6} of gravity since no spin is consider here and we work with teleparallel theory in vacuum. The alternative interpretation here of treating the electrodynamics here as an electromagnetic effective geometry was actually proposed by Visser recently \cite{5} either in the context of HO linear electrodynamics or in Novello's nonlinear electrodynamics \cite{5}. The effective torsion here is analogous to the electric field in much the similar way as the acoustic torsion \cite{7} recently introduced in the literature is analogous to the rotational of the vortex hydrodynamical flow in superfluids \cite{8,9}. Recently we also have show that an effective torsion \cite{10} appears in non-Riemannian geometrical approach to the Lense-Thirring rotational of superfluids. It is also important to stress that our theory is not a Einstein-Maxwell type theory of e.m on a gravitational general relativistic background but actually it represents a true geometrical unification of the gravitational and the e.m fields. The paper is organised as follows: In section 2, the teleparallel $T_{4}$ theory of effective electrodynamice is proposed, in section 3 the Riemannian electrodynamics is developed and finally in section 4 conclusions and discussions are undertaken. 
\section{Teleparallel effective geometry in electrodynamics}
In the Einstein-Cartan spirit, teleparallel theories have been used in the past not as a theory of the role of spin in gravity \cite{11} but rather as a theory where torsion could be used in some way to be related to the e.m field tensor. Our metric as we shall see can be consider as a much simplified form of the Finsler metric which geodesics represent the motion of charged particles and given by \cite{12}
\begin{equation}
ds^{2}= [(g_{ik}dx^{i}dx^{k})^{\frac{1}{2}}+\frac{e}{m}{A_{i}dx^{i}}]^{2}
\label{1}
\end{equation}
where $(i,j=0,1,2)$ and the $A^{i}$ are the e.m potentials that here we split as ${\phi}={\phi}(t,x)$ as the electric potential and $A_{x}(t,x)$ as part of the magnetic vector potential $\vec{A}$. The metric consider here which generates the effective $(2+1)-dimensional$ electrodynamics is given by
\begin{equation}
ds^{2}= (1-{\phi}^{2})dt^{2}-(1-{A_{x}}^{2})dx^{2}-dy^{2}
\label{2}
\end{equation}
Note that this metric can be obtained as a metric perturbation of the Minkowskian $(2+1)$ dimensional metric $ds^{2}=dt^{2}-dx^{2}-dy^{2}$ where the metric perturbation components  are given by $h_{00}=-{\phi}^{2}$ , $h_{xx}=-{A_{x}}^{2}$ and $h_{tx}=-{\phi}A_{x}$, others zero, and the whole spacetime effective metric would be $g_{ik}={\eta}_{ik}+h_{ik}$ where ${\eta}_{ik}$ represents the $M_{4}$ Minkowski spacetime metric. Before we make use of Cartan calculus \cite{13} of differential forms to compute the effective Cartan torsion in terms of the electric field, we express the metric (\ref{2}) in the form  
\begin{equation}
ds^{2}=({\omega}^{0})^{2}- ({\omega}^{1})^{2}-({\omega}^{2})^{2}- ({\omega}^{3})^{2}
\label{3}
\end{equation}
where the basis frame one-form ${\omega}^{a}$ where $a=0,1,2,3$ , are given by
\begin{equation}
{\omega}^{0}=dt
\label{4}
\end{equation}
\begin{equation}
{\omega}^{1}=dx
\label{5}
\end{equation}
\begin{equation}
{\omega}^{2}=dy
\label{6}
\end{equation}
\begin{equation}
{\omega}^{3}={\phi}dt + A_{x}dx
\label{7}
\end{equation}
Note that although we use four indices in differential forms the spacetime is $(2+1)-dimensional$. A similar metric has been used by Letelier \cite{14} in the context of the teleparallel spacetime defects. This metric has been recently generalized to Riemann-Cartan spacetime \cite{15}. By performing the exterior derivatives of the basis one-forms one obtains 
\begin{equation}
T^{3}=d{\omega}^{3}= {T^{3}}_{tx}dt{\wedge}dx= [-{\partial}_{x}{\phi}+{\partial}_{t}A_{x}]dt{\wedge}dx 
\label{8}
\end{equation}
where the symbol ${\wedge}$ means the exterior product of differential forms \cite{16} and use has been made of the Cartan's first structure equation 
\begin{equation}
T^{a}=d{\omega}^{a}+{{\omega}^{a}}_{b}{\wedge}{\omega}^{b}
\label{9}
\end{equation}
where $T^{a}$ represents the Cartan torsion two-form while ${{\omega}^{a}}$ stands for the connection one-form. We also use the choice of the orthonormal frame basis where ${{\omega}^{a}}_{b}$ vanishes which by the second Cartan's structure equation
\begin{equation}
{R^{a}}_{b}={R^{a}}_{bcd}{\omega}^{c}{\wedge}{\omega}^{d}=d{{\omega}^{a}}_{b}+{{\omega}^{a}}_{e}{\wedge}{{\omega}^{e}}_{b}
\label{10}
\end{equation}
implies that the teleparallel constraint of the vanishing of the Riemann-Cartan tensor ${R^{a}}_{bcd}$ is obeyed. Note that the expression (\ref{9}) allows us to identify Cartan effective torsion with the effective electric field, as
\begin{equation}
E_{x}(t,x)= {T^{3}}_{tx}= [-{\partial}_{x}{\phi}+{\partial}_{t}A_{x}] 
\label{11}
\end{equation}
This identification is possible since the RHS of equation (\ref{11}) represents in three dimensions the electric vector field definition 
\begin{equation}
\vec{E}= -{\nabla}{\phi}+ \frac{{\partial}\vec{A}}{{\partial}t}
\label{12}
\end{equation}
as long as ${\phi}$ and $\vec{A}$ are respectively identified with the corresponding electric and magnetic potentials. Thus we may conclude that the only nonvanishing torsion component can be considered as an effective torsion in this non-Riemannian framework of electrodynamics. 
\section{Riemannian electrodynamics from an effective static metric}
Since the teleparallelism is usually seen as a way of expressing the Riemannian Einsteinian gravity (general relativity) theory in terms 
of torsion , in the this section we shall address the particular case of a static effective metric obtained from the line element (\ref{2}) in the case the magnetic potential vanishes, or in the absence of magnetic fields where the electric field turns to be an electrostatic field. We could call this an effective static eletromagnetic  line element  
\begin{equation}
ds^{2}= (1-{\phi}^{2})dt^{2}-dx^{2}-dy^{2}
\label{13}
\end{equation}
This simple metric yields the following components to Riemann, Ricci and Einstein tensors
\begin{equation}
R_{0101}= \frac{[({\phi}-{\phi}^{3}){\partial}_{x}E_{x}-{E_{x}}^{2}]}{(-1+{\phi}^{2})}
\label{14}
\end{equation}
\begin{equation}
R_{11}= \frac{R_{00}}{(-1+{\phi}^{2})}=\frac{[({\phi}-{\phi}^{3}){\partial}_{x}E_{x}-{E_{x}}^{2}]}{(-1+{\phi}^{2})^{2}}
\label{15}
\end{equation}
\begin{equation}
G_{22}= G_{33}=  \frac{[({\phi}-{\phi}^{3}){\partial}_{x}E_{x}-{E_{x}}^{2}]}{(-1+{\phi}^{2})}
\label{16}
\end{equation}
while the Ricci scalar is
\begin{equation}
R = -2\frac{[({\phi}-{\phi}^{3}){\partial}_{x}E_{x}-{E_{x}}^{2}]}{(-1+{\phi}^{2})}
\label{17}
\end{equation}
From there Riemannian curvature expressions one notes that the Einstein vacuum equations 
\begin{equation}
R_{ab}=0 
\label{18}
\end{equation}
implies the following equation
\begin{equation}
{[({\phi}-{\phi}^{3}){\partial}_{x}E_{x}-{E_{x}}^{2}]}{(-1+{\phi}^{2})}=0
\label{19}
\end{equation}
From this expression we notice that in the linear electric field approximation ${E_{x}}^{2}$ we have that 
\begin{equation}
{({\phi}-{\phi}^{3})}{\partial}_{x}E_{x}=0
\label{20}
\end{equation}
which is equivalent to the Coulomb law of effective electrostatic ${\partial}_{x}E_{x}=0$ or in $3-dimensions$ ${\nabla}.\vec{E}=0$. But note that substitution of this Coulomb law into the curvature expression (\ref{14}) yields that the curvature vanishes and we are left with the a flat spacetime effective electrostatic ,while if one considers that the the Einstein vacum equation is not obeyed here the Coulomb law becomes
\begin{equation}
R_{0101}= \frac{({\phi}-{\phi}^{3}){\partial}_{x}E_{x}}{(-1+{\phi}^{2})}
\label{21}
\end{equation}
Note that in this case the Riemann curvature tensor would be proportional to a charge density and we are left with a linear electrodynamics in the electric field but where the charge density is generated by the nonlinear electrostatic potentials. Note that for a Coulombian electrostatic field with bound charges ,the asymptotic behaviour of the electrostatic field would give that ${\phi}^{3}<<{\phi}$ and (\ref{21}) can be approximated to
\begin{equation}
R_{0101}= \frac{{\phi}{\partial}_{x}E_{x}}{(-1+{\phi}^{2})}
\label{22}
\end{equation}
Note that in this case the horizon would imply a divergence in the Riemann tensor. This as in Schwarzschild case does not imply a true singularity, like in astrophysical and optical black holes \cite{17} one has to compute the Krestschmann curvature invariant
\begin{equation}
R_{01tx}R^{01tx}= -\frac{{\phi}^{2}{\partial}_{x}E_{x}}{({1-3{\phi}^{2}})}=\frac{{\partial}_{x}E_{x}}{2}< \infty
\label{23}
\end{equation}
and we note that there is no physical singularity in the effective electrodynamic black hole horizon as in Schwarzschild black hole. Here we use the fact that at the effective black hole horizon ${\phi}=1$. The reader can easily compute the inverse case of magnetic metric where the electrostatic potential vanishes and we are left with a magnetostatic effective metric. The effective black hole considered here could be called an analog electrostatic black hole. Analog black holes in flowing dielectrics have been previously obtained by Novello et al \cite{3} in the context of nonlinear electrodynamics.

\newpage
\section{Discussions and Conclusions}
We present here a non-Riemannian teleparallel framework for effective electrodynamics. A analog electrostatic Riemannian effective black hole has been obtained on the linear electrodynamics similar to the one obtained by Novello et al \cite{3} in the context of flowing dielectrics in Riemannian effective spacetime. Future perspectives may involve the generalization of these ideas to include matter in an effective way and compare this sort of post-metric e.m theory with the pre-metric electrodynamics.                                                 

\end{document}